\documentclass[11pt]{article}
\usepackage{graphicx}
\usepackage{url}
\usepackage{amsmath}
\usepackage{amssymb}
\usepackage[utf8]{inputenc}

\usepackage{ifpdf}
\ifpdf
\usepackage[%
  pdftitle={},%
  pdfauthor={},%
  pdfstartview=FitH,%
  bookmarks=false,%
  bookmarksopen=false,%
  breaklinks=true,%
  colorlinks=true,%
  linkcolor=blue,anchorcolor=blue,%
  citecolor=blue,filecolor=blue,%
  menucolor=black,pagecolor=blue,%
  urlcolor=blue]{hyperref}
\else
\usepackage[%
  breaklinks=true,%
  colorlinks=true,%
  linkcolor=blue,anchorcolor=blue,%
  citecolor=blue,filecolor=blue,%
  menucolor=blue,pagecolor=blue,%
  urlcolor=blue]{hyperref}
\fi
\usepackage{bm}

\date{}

\renewcommand{\title}[1]{\begin{flushleft}{\bf \Large #1}\end{flushleft}}

\begin{document}
\thispagestyle{empty}


\begin{center}
\Large{Estimating effective infection fatality rates during the course of the COVID-19 pandemic in Germany}
\end{center}

\vspace{.1cm}

\begin{center}
{\large
Christian Staerk\footnote[1]{ \textit{Address for
correspondence:} Dr.~Christian Staerk, Department of Medical Biometry, Informatics and Epidemiology, Faculty of Medicine, University of Bonn, Venusberg-Campus 1, 53127 Bonn.\\ Email:
\href{mailto:christian.staerk$at$imbie.uni-bonn.de}{christian.staerk@imbie.uni-bonn.de}}, Tobias Wistuba, Andreas Mayr  \\[2mm]
\vspace{.25cm}
\textit{Working Group Statistical Methods in Epidemiology\\ Department of Medical Biometry, Informatics and Epidemiology, Faculty of Medicine, University of Bonn, Germany} \\[2mm]
}
\end{center}


\begin{abstract} 
\noindent \textbf{Background:} The infection fatality rate (IFR) of the Coronavirus Disease 2019 (COVID-19) is one of the most discussed figures in the context of this pandemic. In contrast to the case fatality rate (CFR), the IFR depends on the total number of infected individuals -- not just on the number of confirmed cases. In order to estimate the IFR, several seroprevalence studies have been or are currently conducted.    \\[1mm]
\noindent \textbf{Methods:} 
\textcolor{black}{Using German COVID-19 surveillance data and age-group specific IFR estimates from multiple international studies, this work investigates time-dependent variations in \textit{effective IFR} over the course of the pandemic.} Three different methods for estimating (effective) IFRs are presented: (a) population-averaged IFRs based on the assumption that the infection risk is independent of age and time, (b) effective IFRs based on the assumption that the age distribution of confirmed cases approximately reflects the age distribution of infected individuals, and (c) effective IFRs accounting for age- and time-dependent dark figures of infections.   \\[1mm]
\noindent  \textbf{Results:}
\textcolor{black}{Effective IFRs in Germany are estimated to vary over time}, as the age distributions of confirmed cases and estimated infections are changing during the course of the pandemic. \textcolor{black}{In particular during the first and second waves of infections in spring and autumn/winter 2020}, there has been a pronounced shift in the age distribution of confirmed cases towards older age groups, resulting in larger effective IFR estimates. \textcolor{black}{The} temporary increase in effective IFR \textcolor{black}{during the first wave} is estimated to be smaller but still remains when adjusting for age- and time-dependent dark figures. A comparison of effective IFRs with observed CFRs indicates that a substantial fraction of the time-dependent variability in observed mortality can be explained by changes in the age distribution of infections. Furthermore, a \textcolor{black}{vanishing} gap between effective IFRs and observed CFRs is apparent \textcolor{black}{after the first infection wave, while a moderately increasing gap can be observed during the second wave.}  \\[1mm] 
\noindent \textbf{Conclusions:} 
As the  mortality of COVID-19 largely increases with age, it is important to take the \textcolor{black}{changing} age distribution of infected individuals into account to determine the effective IFR. Further research is warranted to obtain timely age-stratified IFR estimates.





\end{abstract}
  
\clearpage

\section{Background}

The ongoing pandemic of the novel coronavirus disease COVID-19 provides enormous global challenges for public health, society and economy. \textcolor{black}{An important figure in the context of this pandemic is the \textit{infection fatality rate (IFR)}, defined by the number of COVID-19 associated deaths divided by the total number of infections. In contrast to the \textit{case fatality rate (CFR)}, the IFR is not only based on the number of confirmed cases and should therefore not be biased by potential drifts in testing policies.} However, as the total number of infections with SARS-CoV-2 is generally unknown, the IFR can only be estimated based on available surveillance and seroprevalence data. \\




Many seroprevalence studies have been conducted worldwide with the aim of estimating the true numbers of infections and resulting IFRs. Also in Germany, several local seroprevalence studies are being conducted (e.g.~\cite{radon2020}); a completed study from the early phase of the pandemic in the high-prevalence region of Gangelt, Heinsberg, reports an estimated population-averaged IFR of 0.41\% (95\% confidence interval [0.33\%; 0.52\%]), based on 8 observed deaths {\color{black}until 20th of April 
\cite{streeck2020}.}  
Overviews of {\color{black}completed studies} from a wide range of countries can for example be found in \cite{meyerowitz2020} and \cite{ioannidis2020}. The meta-analysis of Meyerowitz-Katz et al. \cite{meyerowitz2020} yields an estimated population-averaged IFR of 0.68\% [0.53\%; 0.82\%], while in the meta-analysis of Ioannidis \cite{ioannidis2020} estimated population-averaged IFRs range from 0.02\% to 0.86\% with a median IFR of 0.26\%.
Although the two meta-analyses differ heavily regarding their results and conclusions, both observe a high heterogeneity in population-averaged IFR estimates among different studies and emphasize the importance of obtaining reliable age-stratified estimates.     \\

More recent meta-analyses \cite{o2020,levin2020} investigate age-specific mortality by estimating infection fatality rates for different age groups. 
As the risk of death from COVID-19 is estimated to increase exponentially with age \cite{levin2020}, it is crucial that the age distribution of infections is taken into account when interpreting estimated ``overall'' (population-averaged) IFRs from different seroprevalence studies in different regions. In fact, a large proportion of the variability in estimated IFRs between studies in different regions may be explained simply by differences in demographics, particularly the age structure of populations. In addition, other factors such as the prevalence of certain comorbidities, the access to intensive medical care and systematic differences in social networks might also contribute to the variability of COVID-19 associated mortality in different regions. \\

However, even when considering only a particular region, the observed mortality of SARS-CoV-2 may vary over time, as the  \textcolor{black}{age distribution of the infected population} may be changing throughout the pandemic, e.g.\ due to different and changing risk behaviours \textcolor{black}{as well as specific preventive measures for high-risk groups.} 
In a recent study regarding ``the foreshadow of a second wave'' in Germany, Linden et al.~\cite{linden2020} estimate the \textit{effective IFR} as a time-dependent measure \textcolor{black}{by considering the infection fatality rate given the changing age distribution of confirmed cases, using age-specific IFR estimates from the meta-analysis of Levin et al.~\cite{levin2020}.} 
The effective IFR is estimated based on the assumption that the distribution of confirmed cases reflects the distribution of true infections \cite{linden2020}. This may not necessarily be the case, as e.g.\ infections with SARS-CoV-2 may be more likely detected in older age groups since these tend to experience a more serious disease course. Furthermore, as the focus of the study by Linden et al.~\cite{linden2020} has been on the analysis of the {\color{black}beginning} of a second wave of infections during late summer 2020, effective IFRs have not been reported for the earlier phase of the pandemic in Germany. \\

Thus, our study aims to investigate \textcolor{black}{time-dependent variations} in effective IFR over the course of the COVID-19 pandemic in Germany, by combining age-specific IFR estimates from multiple studies with publicly available German surveillance data. We compare \textcolor{black}{estimated effective IFRs} based on the age distribution of confirmed cases with estimates derived from the age distribution of estimated infections, obtained through estimated age- and time-dependent dark figures. Results are presented based on age-specific IFR estimates from four different studies, illustrating the remaining uncertainty regarding age-specific {\color{black}mortality.} 

\section{Methods}

\textcolor{black}{We use the German COVID-19 surveillance data provided by 
the Robert Koch Institute \cite{RKI_data}, containing information on date of disease onset (or date of confirmation of SARS-CoV-2 infection if disease onset unknown) and information on deaths associated with COVID-19 for individual confirmed cases.} 
Data on age of confirmed cases and deaths are available for the following age groups \(A= \{\text{0-4}, \text{5-14}, \text{15-34}, \text{35-59}, \text{60-79}, \text{80+}\}\). \\

\begin{figure} [t!]
\includegraphics[scale = 0.55]{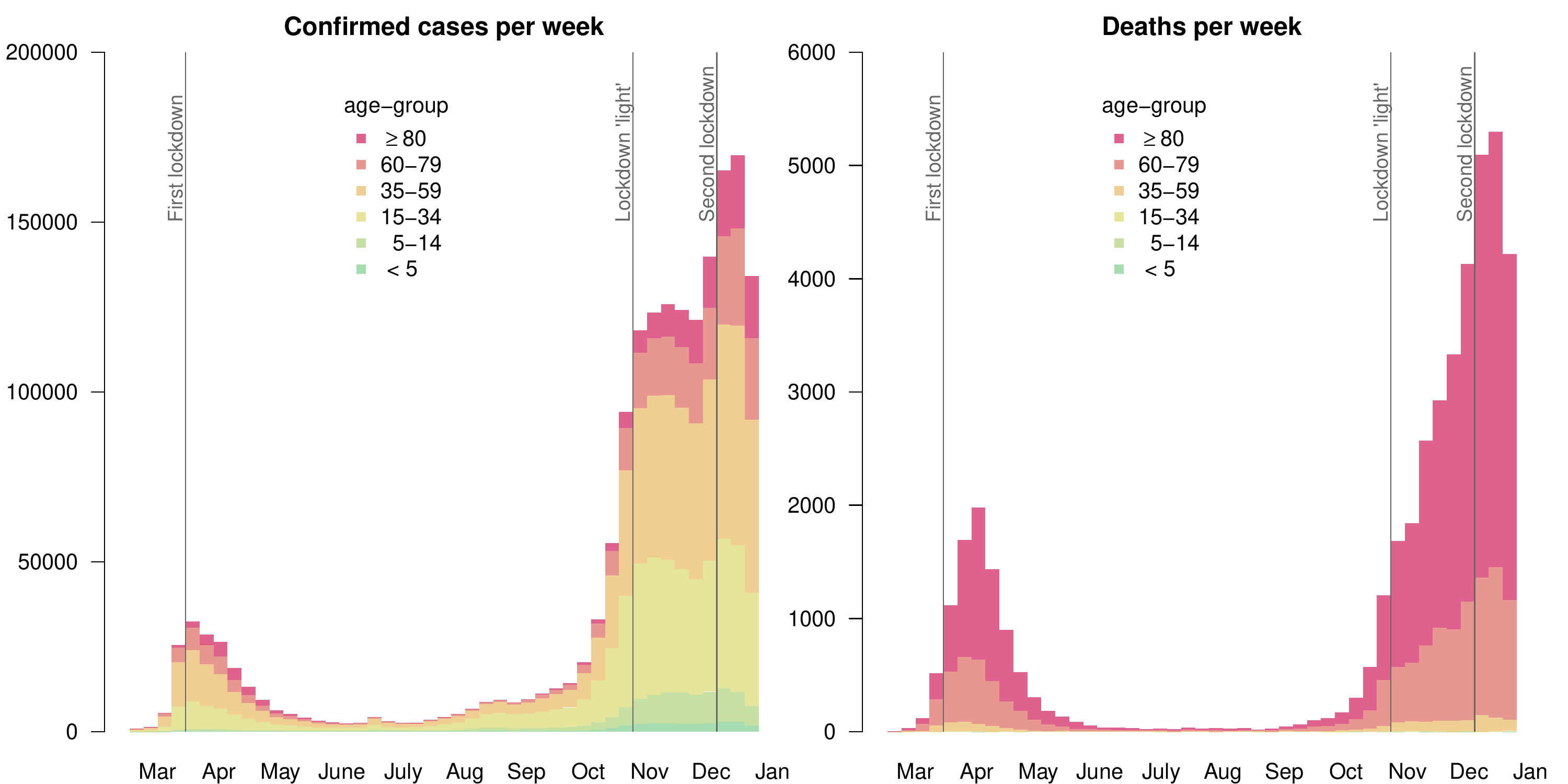}
\caption{\label{fig:weekly_case_death}\small Absolute numbers of weekly confirmed cases  \(C_{a,t}\) (left plot) and deaths \(D_{a,t}\) (right plot) for the different age groups in Germany \textcolor{black}{for the year 2020 (data as of January 21, 2021 \cite{RKI_data})}. In both graphics, the time points $t$ refer to the respective weeks of disease onset (or of confirmation of \textcolor{black}{infection if disease onset unknown}). \textcolor{black}{Additionally, vertical lines indicate the introduction of major mitigation measures in Germany (see also Table~\ref{tab:measures}).}}
\end{figure}

\textcolor{black}{We} consider cumulative data for each calendar week, so that potential weekday-specific \textcolor{black}{fluctuations are eliminated}. Let \(C_{a,t}\) denote the number of confirmed cases for age group \(a\in A\) in calendar week \(t\) and let \(C_{t} = \sum_{a\in A} C_{a,t}\) denote the total number of confirmed cases in week \(t\). Similarly, let \(I_{a,t}\) and \(I_{t}\) denote the number of true infections and let \(D_{a,t}\) and \(D_{t}\) denote the number of deaths, for age group \(a\) and week \(t\). Note that deaths typically occur weeks after the onset of symptoms from COVID-19 with an estimated \textcolor{black}{average interval} of 16 days \cite{khalili2020}, while infections with SARS-CoV-2 occur several days prior to manifestation of disease with an estimated median incubation period of 5 days \cite{lauer2020}. However, here the time point \(t\) in \(C_{a,t}\), \(I_{a,t}\) and \(D_{a,t}\) always refers to the same week where the \textit{infection} has been \textcolor{black}{manifested or confirmed}.  
\textcolor{black}{The course of weekly confirmed cases and deaths for the different age groups in Germany is depicted in Figure~\ref{fig:weekly_case_death}.} \\

The observed CFR in week \(t\) is defined by the number of deaths \(D_t\) (resulting from infections in week \(t\)) divided by the number of confirmed cases \(C_t\) in week \(t\), i.e.  \(\text{CFR}_{t} = \frac{D_t}{C_t}\). On the other hand, the effective IFR in week \(t\) is defined based on the total number of infections, i.e.\  
\(\text{IFR}_{\text{eff},t} = \frac{D_t}{I_t}\). As the number of infections \(I_t\) is unknown, we estimate the weekly effective IFR by taking into account the time-dependent distribution of infections in the different age groups \(a\in A\) as well as age-specific IFR estimates from four different studies. 
In particular, we consider one modelling study \cite{verity2020} estimating age-specific IFR for China based on individual-case data from the early phase of the pandemic (until 25th of February), one seroprevalence study \cite{perez2020} in Geneva, Switzerland (until 1st of June) specifically designed for age-stratified estimation of IFR, as well as two recent meta-analyses \cite{o2020, levin2020} combining the evidence from multiple seroprevalence studies worldwide. 
As the age groups in the four sources of estimated IFR slightly differ from the age groups \(A\) considered in the used surveillance data, age-specific IFR estimates are adjusted to match with the age groups \(A\) via weighted averaging of estimates, taking into account the age structure of the  German population (based on data from \cite{Destatis}). \\

Let \(\widehat{\text{IFR}}_a^{(i)}\) denote the resulting estimated IFR for age group \(a\) from study $i$, with the index $i$ referring to one of the four literature sources (\(i \in\{\text{O'Driscoll} \cite{o2020}, \text{Verity} \cite{verity2020}, \text{Perez-Saez} \cite{perez2020}, \text{Levin} \cite{levin2020}\}\)). The true effective \(\text{IFR}_{\text{eff},t}\) can be estimated by a weighted average of age-specific IFR estimates, i.e.\ 
\begin{equation} \label{eq:estimator}
    \widehat{\text{IFR}}_{\text{eff},t}^{(i)} =  \sum_{a\in A} \hat{\omega}_{a,t} \cdot \widehat{\text{IFR}}_a^{(i)} \,,
\end{equation}
where \(\hat{\omega}_{a,t}\) denotes an estimator for the fraction of infections \(I_{a,t}/I_{t}\) in age group \(a\) in week \(t\). Note that estimator \eqref{eq:estimator} for the effective \(\text{IFR}_{\text{eff},t}\) is based on the crucial assumption that age-dependent infection fatality rates \(\text{IFR}_a\) do not change over the course of the pandemic and that the estimates of \(\text{IFR}_a\) from the four international studies are applicable to Germany.  \\

In the following, we consider three different estimators for the fraction of infections \(\hat{\omega}_{a,t}\), which are derived under different assumptions regarding the distribution of infections. 

\begin{enumerate}

\item[(a)] Under the theoretical assumption that the risk of infection is independent of age and time (compare also \cite{linden2020}), 
the effective IFR -- denoted by \(\overline{\text{IFR}}_{\text{DE}}^{(i)}\) -- is constant over time and estimated by equation \eqref{eq:estimator} with time-independent weights
\begin{equation}
    \hat{\omega}_{a,t} =  \hat{\omega}_{a} = \frac{P_{a}}{P} \,,
\end{equation}
where the population numbers \(P_a\) in age groups \(a\in A\) and the total population number \(P=\sum_{a\in A} P_a\) of Germany are regarded as constant over time. 

\item[(b)] In practice, the (non-uniform) age distribution of infections is likely to be changing over the course of the pandemic. Under the assumption that the distribution of confirmed cases approximately reflects the distribution of true infections in the different age groups, i.e. 
\begin{equation} \label{eq:assumption1}
   \frac{C_{a,t}}{C_t} \approx \frac{I_{a,t}}{I_t} \,,
\end{equation}
one can estimate the fraction of unknown infections in age group \(a\) in week \(t\) by the corresponding fraction of confirmed cases
\begin{equation} \label{eq:estimator_omega1}
  \hat{\omega}_{a,t} = \frac{C_{a,t}}{C_t} \,.
\end{equation}
Assumption \eqref{eq:assumption1} and the resulting estimator \eqref{eq:estimator_omega1} correspond also to the effective IFR defined in \cite{linden2020}. 
Note that assumption (\ref{eq:assumption1}) implies that dark figures of undetected infections are approximately independent of age. 

\item[(c)]  Without specific assumptions as in (a) and (b), the number of infections \(I_{a,t}\) can be alternatively estimated by considering age- and time-dependent dark figures via 
\begin{align} \label{eq:assumption2}
   &\hat{I}_{a,t}^{(i)} = \hat{f}_{a,t}^{(i)} \cdot C_{a,t}\,, \nonumber 
   \\ &~~~~\text{with}~ \hat{f}_{a,t}^{(i)} = \frac{\text{CFR}_{a,t}}{\widehat{\text{IFR}}_a^{(i)}} = \frac{ \frac{D_{a,t}}{C_{a,t}}}{ \widehat{\text{IFR}}_a^{(i)} }   \,,
\end{align}
where \(\hat{f}_{a,t}^{(i)}\) denotes the estimated factor for the dark figure in age group \(a\) in week \(t\) based on study \(i\). Thus, an alternative estimator for the fraction of all infections in age group \(a\) in week \(t\) is given by 
\begin{equation} \label{eq:estimator2}
  \hat{\omega}_{a,t}^{(i)} = \frac{\hat{I}^{(i)}_{a,t}}{\sum_{a'\in A}\hat{I}^{(i)}_{a',t}} = \frac{\hat{f}_{a,t}^{(i)} \cdot C_{a,t}}{ \sum_{a'\in A} \hat{f}_{a',t}^{(i)} \cdot C_{a',t} }  \,.
\end{equation} 
As the number of true infections \textcolor{black}{is at least as high as the number of confirmed cases}  (\(I_{a,t} \geq C_{a,t}\)), the corresponding factor \(f_{a,t} = I_{a,t}\,/\,C_{a,t}\)  for the dark figure should be lower bounded by 1. Thus, in the following we use the estimator \(  \hat{f}_{a,t}^{(i)} = \max\{1, \text{CFR}_{a,t} \,/\, \widehat{\text{IFR}}_a^{(i)} \}\). Due to relatively small numbers of observed deaths in younger age groups, we combine the age groups \(a\in\{\text{0-4}, \text{5-14}, \text{15-34}, \text{35-59}\}\) yielding joint estimates \(\hat{f}_{0-59,t}^{(i)}\) of time-dependent dark figures for ages 0 to 59. To further stabilize the \textcolor{black}{procedure}, we estimate monthly (instead of weekly) dark figures based on age-specific CFRs observed for each month.   
\end{enumerate}
In this study we hence compare three different estimators for the effective IFR, each depending on different assumptions. Estimator (a) leads to a time-constant effective \(\overline{\text{IFR}}_{\text{DE}}^{(i)}\), while estimators (b) and (c) actually take into account the {\color{black}changing} age distribution of confirmed cases and estimated infections, respectively. All three estimators depend on the four available age-specific estimates $\widehat{\text{IFR}}_a^{(i)}$ from the literature (Table~\ref{tab:IFR}).

\section{Results}

 
\textcolor{black}{The risk of death among persons infected with SARS-CoV-2 is estimated to increase substantially with increasing age by each of the four considered studies 
(Table~\ref{tab:IFR}), which is also supported by the number of observed deaths $D_{a,t}$ per age group in Germany (see Figure~\ref{fig:weekly_case_death}).
However, estimates from the literature show larger discrepancies; as for example in age group 80+, the IFR estimate from \cite{perez2020} is given by 5.60\% [4.30\%; 7.40\%], while the corresponding IFR estimate from \cite{levin2020} is as large as 15.61\% [12.20\%; 19.50\%].} On the other hand, for the age group 60-79 the IFR estimate from \cite{o2020} is approximately 1\%, while the other studies yield larger estimates for this age group ranging from 2.49\% in \cite{levin2020} to 3.89\% in \cite{perez2020}.
Furthermore, Table~\ref{tab:IFR} gives estimates of resulting population-averaged infection fatality rates \(\overline{\text{IFR}}_{\text{DE}}^{(i)}\) for Germany, which are derived under the assumption that the risk of infection with SARS-CoV-2 is independent of age and time (see assumption (a)). Population-averaged estimates \(\overline{\text{IFR}}_{\text{DE}}^{(i)}\) for Germany range from 0.756\% [0.717\%; 0.796\%] by \cite{o2020} to 1.687\% [1.407\%; 2.139\%] by \cite{levin2020}, reflecting the uncertainty regarding age-specific IFR.   \\

\begin{figure}[t!] 
\includegraphics[scale = 0.65]{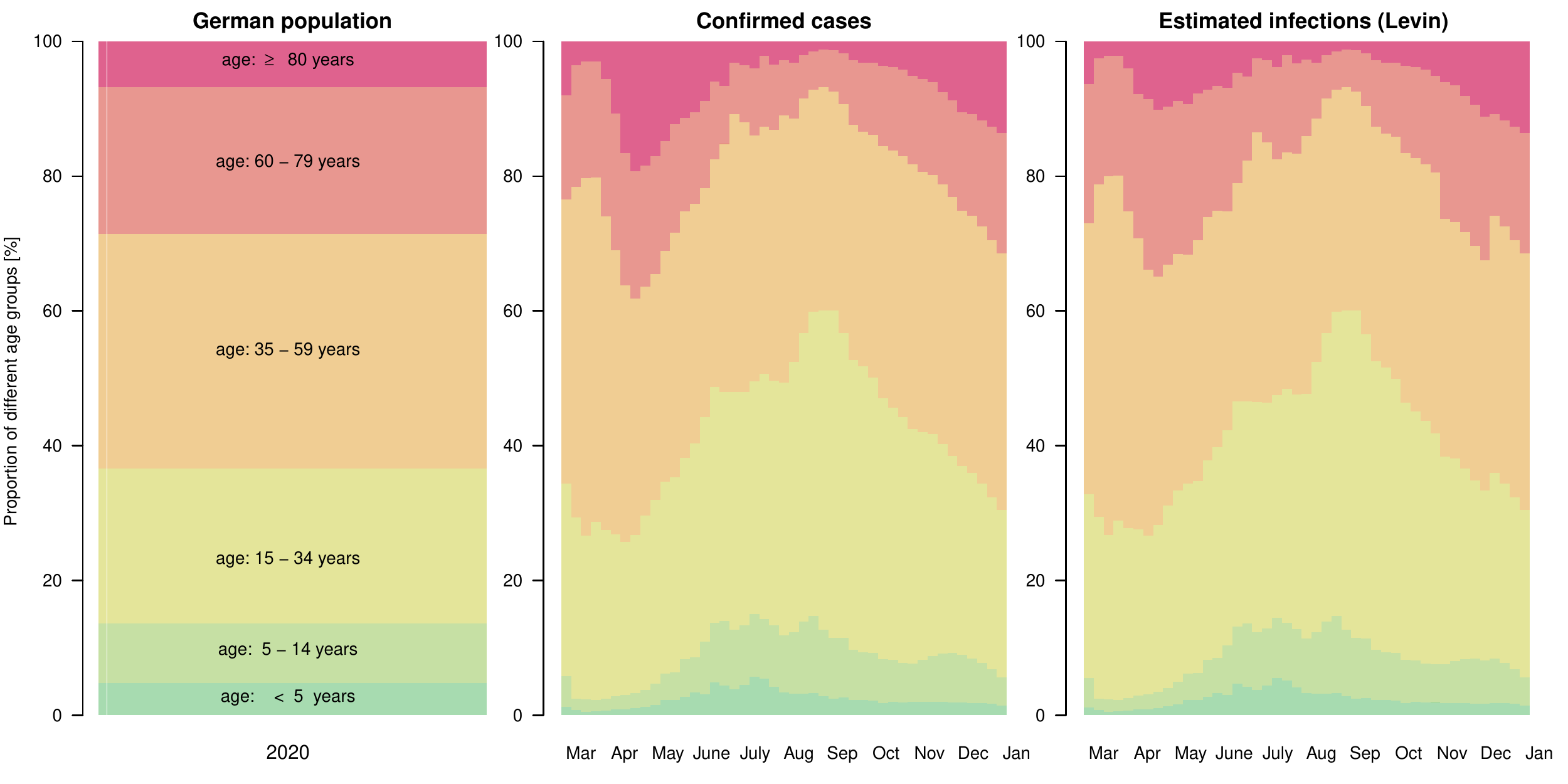}
\caption{\label{fig:weekly_case_inf} \small Age distributions of the \textcolor{black}{general} German population (left plot), of weekly confirmed cases (central plot) and of estimated weekly numbers of infections (right plot) based on the age-specific IFR estimates by Levin et al. \cite{levin2020}.}
\end{figure}

\begin{table}[t!]
\caption{\small Age-group specific estimates \(\widehat{\text{IFR}}_a^{(i)}\) as well as  population-averaged estimates \(\overline{\text{IFR}}_{\text{DE}}^{(i)}\) for Germany under age-independent infection risk,  based on studies \(i \in\{\text{O'Driscoll}\cite{o2020}, \text{Verity}\cite{verity2020}, \text{Perez-Saez} \cite{perez2020}, \text{Levin} \cite{levin2020}\}\). IFR estimates are given in percentages (with 95\% confidence intervals in brackets).\label{tab:IFR}} 
\small
\begin{center}
 \begin{tabular}{r|cccc} 
  \hline
 Age group  & O'Driscoll \cite{o2020} & Verity \cite{verity2020} & Perez-Saez \cite{perez2020} & Levin \cite{levin2020} \\ 
  \hline
0-4 & 0.002 [0.001; 0.002] & 0.002 [0.000; 0.025] & 0.002 [0.000; 0.019] & 0.001 [0.001; 0.001] \\ 
  5-14 & 0.000 [0.000; 0.000] & 0.004 [0.001; 0.037] & 0.001 [0.000; 0.011] & 0.002 [0.001; 0.003] \\ 
  15-34 & 0.009 [0.007; 0.010] & 0.041 [0.019; 0.110] & 0.007 [0.003; 0.013] & 0.016 [0.014; 0.020] \\ 
  35-59 & 0.122 [0.115; 0.128] & 0.349 [0.194; 0.743] & 0.070 [0.047; 0.097] & 0.226 [0.212; 0.276] \\ 
  60-79 & 0.992 [0.942; 1.045] & 2.913 [1.670; 5.793] & 3.892 [2.985; 5.145] & 2.491 [2.294; 3.266] \\ 
  80+ & 7.274 [6.909; 7.656] & 7.800 [3.800; 13.30] & 5.600 [4.300; 7.400] & 15.61 [12.20; 19.50] \\ \hline
   \(\overline{\text{IFR}}_{\text{DE}}^{(i)}\) & 0.756 [0.717; 0.796] & 1.296 [0.694; 2.453] & 1.254 [0.959; 1.661] & 1.687 [1.407; 2.139] \\ 
   \hline
\end{tabular}
\end{center}
\end{table}
The estimated population-averaged infection fatality rates \(\overline{\text{IFR}}_{\text{DE}}^{(i)}\), based on different age-specific IFR estimates, can be interpreted as reference mortality figures for the \textcolor{black}{\textit{general}} German population in order to compare them to other countries. They have a rather theoretical meaning as they do not reflect the actual age distribution of \textcolor{black}{the \textit{infected} population.}  Figure~\ref{fig:weekly_case_inf} depicts the \textcolor{black}{changing age distribution of weekly confirmed cases} (central plot) in comparison to the age distribution of the \textcolor{black}{general population} (left plot). 
It can be observed that the age distribution of confirmed cases shifted considerably towards older age groups during the first wave in Germany in March and April 2020. 
\textcolor{black}{During summer} with a relatively low incidence of COVID-19, confirmed cases were predominantly observed in younger age groups. However, since September 2020, percentages of confirmed cases among the elderly have been \textcolor{black}{continuously} rising again. \textcolor{black}{In the end of December 2020, the age distribution of confirmed cases is remarkably similar to the distribution of confirmed cases in April during the first wave of infections.}  \\

\begin{figure}[t!]
\includegraphics[scale = 0.65]{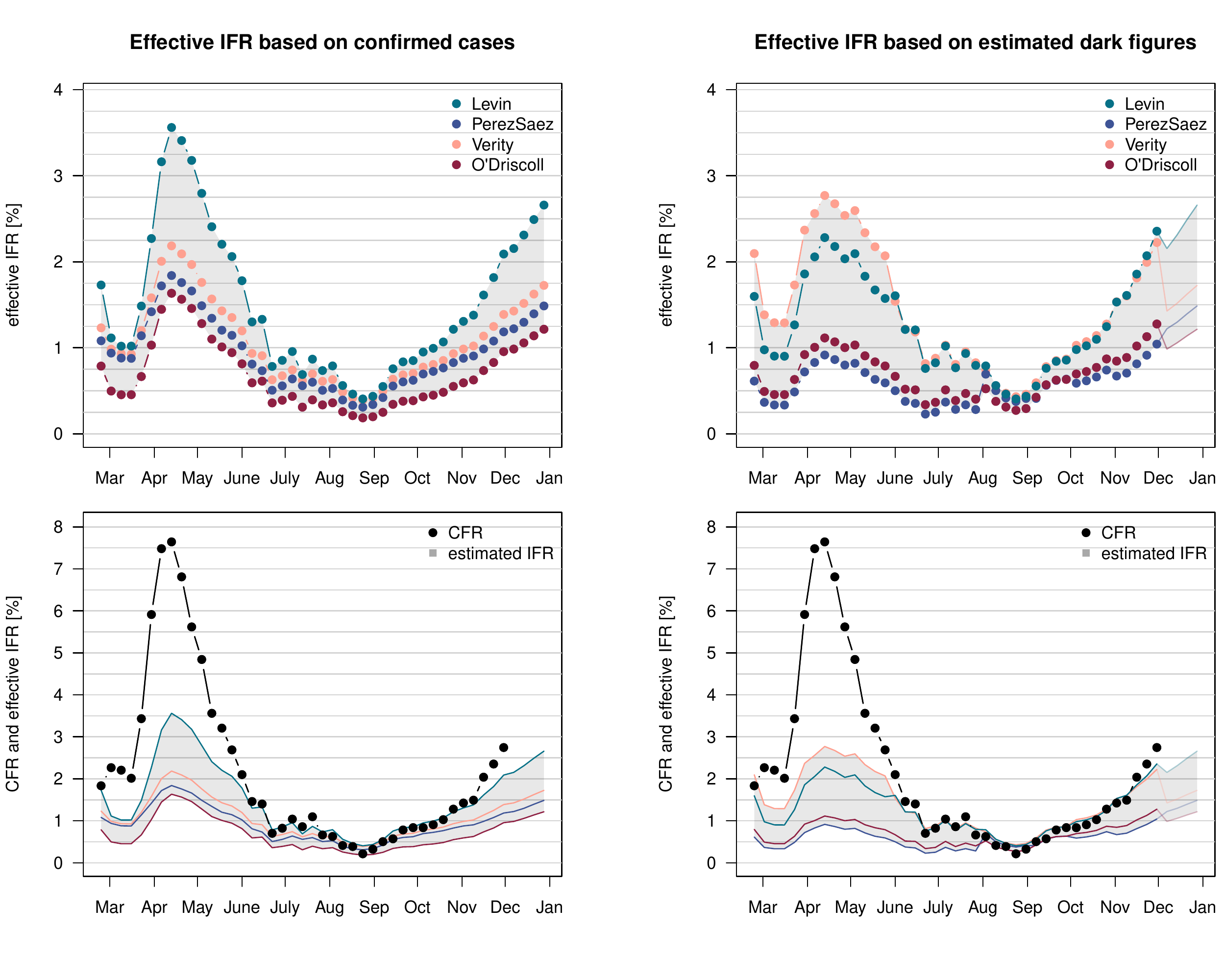}
\caption{\label{fig:effective_IFR} \small \textcolor{black}{Estimated} (weekly) effective infection fatality rates $ \widehat{\text{IFR}}_{\text{eff},t}^{(i)} $ \textcolor{black}{in Germany for the year 2020,} based on four different age-specific IFR estimates (\(i\)). The left plots refer to the estimates with method (b) based on the age distribution of confirmed cases, while the plots on the right refer to the estimates with method (c) based on the age distribution of infections via  estimated age- and time-dependent dark figures. The lower plots additionally display \textcolor{black}{observed CFRs} in Germany. Due to the expected time lag between infections and deaths, observed CFRs are only shown until the end of \textcolor{black}{November}; similarly, effective IFR estimates with method (c) are only accounted for estimated dark figures (based on observed CFRs) until the end of \textcolor{black}{November}.}
\end{figure}

The described trend in the distribution of confirmed cases over time is directly reflected in the corresponding \textcolor{black}{development} of estimated effective IFR (based on method (b)). The left side in Figure~\ref{fig:effective_IFR} shows that the estimated effective IFR sharply increases from values between 0.5\% and 1\% in March to values between 1.5\% and 3.5\% in April. After this peak, the estimated effective IFR has been declining to values between 0.2\% and 0.5\% in the end of August, corresponding to a relatively young age distribution of confirmed cases. This observation may be partly explained by an increased mobility of younger age groups during the summer holiday period. Since September 2020, as the distribution of confirmed cases has been shifting more towards older age groups, effective IFR estimates \textcolor{black}{have been rising again up to similar levels as in the peak during the first wave of infections.} 
This indicates that with larger SARS-CoV-2 incidences (see Figure~\ref{fig:weekly_case_death}) it may become increasingly difficult to effectively protect vulnerable risk groups and to prevent the spread of the virus from younger to older age groups (see also \cite{linden2020}). \\

As the age distribution of \textit{confirmed cases} may not generally reflect the age distribution of \textit{true infections}, in a further analysis we account for age- and time-dependent dark figures (see method~(c)). The right hand side of Figure~\ref{fig:weekly_case_inf} depicts the \textcolor{black}{development} of estimated true infections based on IFR estimates from Levin et al.~\cite{levin2020}. It can be seen that the \textcolor{black}{development} of estimated infections is similar in shape to the observed \textcolor{black}{development} of confirmed cases. However, in particular following the high phase of the first wave of infections in April (compare Figure~\ref{fig:weekly_case_death}), the estimated distribution of infections is shifted towards younger age groups in comparison to the distribution of confirmed cases. This shift results from dark figures of infections which are estimated to be larger in younger age groups in comparison to the age group 80+ during this particular time. A plausible explanation for this observation might be that in times of limited testing capacities, preferential testing of individuals in age group 80+ has been more pronounced, as these patients are more likely to show (severe) symptoms from COVID-19 requiring medical intervention. \textcolor{black}{Similar} effects on estimated infections during the first wave are also observed when using age-specific IFR estimates from \textcolor{black}{O’Driscoll et al.~\cite{o2020} and Perez-Saez et al.~\cite{perez2020}, whereas numbers of infections in age group 80+ are estimated to be comparatively larger based on Verity et al.~\cite{verity2020} 
(detailed results on estimated infections not shown).} \textcolor{black}{During summer 2020}, there seems to be a close alignment of estimated infections with confirmed cases, as age-dependent factors for dark figures are estimated to be close to 1 (and would have partly been even below 1). This may indicate that a large proportion of infections has been detected with \textcolor{black}{the implemented testing policies during the summer period.} \\ 

The right hand side of Figure~\ref{fig:effective_IFR} depicts the resulting \textcolor{black}{development} of estimated effective IFR when accounting for age- and time-dependent dark figures. It can be seen that the adjustment for \textcolor{black}{dark figures} has a particular effect during the first wave of the pandemic in Germany, where estimated effective IFRs tend to be smaller in comparison to the unadjusted estimates based on confirmed cases (compare to left hand side of Figure~\ref{fig:effective_IFR}). However, even when adjusting for age-dependent dark figures, there still remains a pronounced increase in estimated effective IFRs during the first wave of \textcolor{black}{infections; this indicates that} the increase in mortality cannot exclusively be explained by preferential testing, but that there has been an \textcolor{black}{actual change} in the age distribution of the \textcolor{black}{infected population. During} summer 2020, the age distribution of estimated infections more closely aligns with the age distribution of confirmed cases and thus the estimates of effective IFR adjusted for dark figures are very similar to the unadjusted estimates. \textcolor{black}{In contrast to the first wave of infections in spring 2020, during the second wave the adjusted estimates are not systematically lower and partly even larger than the unadjusted ones.} \\ 

Figure~\ref{fig:effective_IFR} also shows the \textcolor{black}{development} of \textit{estimated effective IFR} in comparison with the \textcolor{black}{development} of \textit{observed CFR} in Germany. It can be seen that trends in observed CFR closely resemble trends in effective IFR estimated based on the age distribution of confirmed cases (as well as true infections). This implies that the {\color{black}age distribution} of infections is a major determinant (and predictor) for the resulting mortality associated with COVID-19. Despite this, it can be observed that the gap between \textcolor{black}{CFR and IFR has been declining after the first wave in Germany}; in fact, observed CFRs in August and September are even lower than some estimates of effective IFR. There may be multiple possible reasons for the decline in CFR, which are independent of the age distribution of infections: The first and probably largest contribution to the observed decline in CFR is the steady and considerable increase in conducted SARS-CoV-2 testing in Germany \cite{RKI_tests}. \textcolor{black}{For example, the number of conducted SARS-CoV-2 tests increased from 586{,}620 in calendar week 31 (end of July) to 1{,}121{,}214 tests in calendar week 35 (end of August) \cite{RKI_14_10}, reflecting a doubling in weekly conducted tests in Germany, partly due to increased testing in the context of the summer holiday season.} 
Another plausible reason for a further decline in observed CFR may be due to improvements in treatment of COVID-19 or other factors leading to a decrease in age-specific IFR \cite{horwitz2020}. \textcolor{black}{During the second wave of infections in autumn/winter 2020, the gap between CFR and IFR tends to be increasing again, but the CFR is still at much lower levels in comparison to the first wave, as dark figures of infections during the second wave are estimated to be smaller than during the first.}  
\clearpage

\textcolor{black}{The course of the pandemic should also be viewed in light of mitigation efforts in Germany (cf.~\cite{ebrahim2020,ebrahim2020b}). Due to the German federal structure there have been specific differences in implemented measures between the 16 federal states, although a uniform procedure has been sought by the federal government and local states. Table~\ref{tab:measures} provides an overview of important mitigation measures which have been applied to most regions of Germany. The effectiveness of interventions during the first infection wave has been investigated by Dehning et al.~\cite{Dehning2020}, concluding that the entirety of measures in context of the ``first lockdown'', implemented in three consecutive weeks  (cf.\ Table~\ref{tab:measures}), effectively reduced the spread of the virus (cf.\ Figure~\ref{fig:weekly_case_death}).} \\

\begin{table}[t!]
\caption{\small \textcolor{black}{Time line of COVID-19 mitigation measures \cite{ebrahim2020,ebrahim2020b} implemented in Germany (cf. Dehning et al. \cite{Dehning2020}). Note that, due to the federal structure of Germany, 
there have been specific differences in implemented measures between the 16 German federal states (not listed here)}. \label{tab:measures}}
\begin{center}
 {\color{black} \small \resizebox{\textwidth}{!}{\begin{tabular}{r|l}
  \hline
 Date  &  Mitigation measures \\ 
  \hline
  10 Mar 2020 & Cancellation of large public events. \\
  16 Mar 2020 & Closures of schools, childcare facilities and non-essential stores. \\
  23 Mar 2020 & \textit{First lockdown}, including strict contact restrictions. \\
  27 Apr 2020 & Beginning of reopening of stores and schools. Mask requirements in stores and public transport. \\
  06 May 2020 & Relaxation of several measures, including less stringent contact restrictions. \\
  16 Jun 2020 & Introduction of German tracing app (``Corona-Warn-App''). \\
  16 Jul 2020 & Specification of ``hotspot strategy'' with targeted local measures in particularly affected districts. \\
  02 Nov 2020 & \textit{Lockdown ``light''}, including stricter contact restrictions and closures of restaurants and leisure facilities.  \\
  16 Dec 2020 & \textit{Second lockdown}, including closures of non-essential stores and switch to distance learning in schools.  
\end{tabular}}}
\end{center}
\end{table}
\textcolor{black}{In addition, our results (Figure~\ref{fig:effective_IFR}) show a time-delayed decline of effective IFR in May 2020 after the incidence peak in March and April, indicating that the spread of the virus to older age groups could be reduced after incidences reached more controllable levels.  During summer 2020, with relaxed mitigating measures still in place, such as the requirement of wearing masks in stores and public transport, incidences of COVID-19 remained relatively low. In this time, there was a particularly young age distribution of cases, resulting in small estimated effective IFRs and a reduced disease burden for high-risk groups. As the second wave of infections in Germany has been ongoing at the time of writing, it is too early for a conclusive evaluation of mitigation measures, including the ``lockdown light'' and the ``second lockdown'' (see Table~\ref{tab:measures} and Figure~\ref{fig:weekly_case_death}); however, our results show that estimated effective IFRs (and observed CFRs) have been continuously rising with increasing numbers of cases until December 2020.} \\

Limitations of this study include that the analysis is based on the assumptions that age-specific IFRs are constant over time and that IFR estimates from the four international studies are applicable to Germany. Furthermore, in this work we have focused on estimating the effective IFR based on the {\color{black}changing} age distribution of infections; however, in practice many other factors may also contribute to the variability in mortality of COVID-19, such as the distribution of different comorbidities as well as the sex distribution of infected individuals. Another limitation of this study is that, due to relatively small numbers of observed deaths in young age groups, monthly dark figures are estimated jointly for the wide age group of 0 to 59 years, even though true dark figures may further differentiate in practice (see e.g.\ \cite{hippich2020} for a recent study of dark figures for children in Germany). Finally, for the estimation of age- and time-dependent dark figures we assumed that there are no systematic biases in reported age-specific deaths, which may not necessarily be the case; for example, Michelozzi et al. \cite{michelozzi2020} investigate the temporal dynamics in excess mortality in Italian cities and observe an underestimation of COVID-19 deaths for older age groups.

\section{Conclusions}

\textcolor{black}{We} have illustrated that the effective IFR in Germany is estimated to vary during the course of the pandemic, as the age distribution of infections is changing over time. In fact, it can be observed that a large fraction of the time-dependent variability in CFR can be explained by {\color{black}changes in} the age distribution of infections. The additionally observed \textcolor{black}{trends} in the gap between the CFR and effective IFR {\color{black}require} further investigation in order to disentangle the contributions of {\color{black}shifts in testing policies} and of other factors that may induce changes in mortality. In particular, obtaining reliable and timely age-specific IFR estimates for Germany is an important issue for further research.     


\small

\bibliographystyle{vancouver}
\bibliography{bibliography}

\end{document}